\documentclass[a4paper]{scrartcl}
\pdfoutput=1
\usepackage[utf8]{inputenc}

\usepackage[T1]{fontenc} 
\usepackage{lmodern} 
\usepackage[sc,osf]{mathpazo} 
\AtBeginDocument{%
  \DeclareFontShape{T1}{pplj}{m}{scit}{<-> ec-qplri-sc}{}
}
\usepackage{myT1classico} 
\DeclareMathAlphabet{\mathsf}{T1}
  {\sfdefault}{m}{n} 
\SetMathAlphabet{\mathsf}{bold}{T1}{\sfdefault}{b}{n} 

\usepackage{amsmath,amssymb,amsthm,mathtools}
\usepackage{tensor}
\usepackage{accents}

\usepackage{enumitem}

\usepackage{accsupp}
\newcommand*{\acronym}[1]{\texorpdfstring{%
    \protect\BeginAccSupp{%
      method=pdfstringdef,%
      ActualText=#1
    }%
    \textsc{\textls[40]{\MakeLowercase{#1}}}%
    \protect\EndAccSupp{}}%
  {#1}%
}

\usepackage[british]{babel}

\usepackage{csquotes}

\usepackage{microtype} 

\usepackage{authblk}

\usepackage{todonotes}

\usepackage[colorlinks]{hyperref}
\usepackage[nameinlink]{cleveref}

\usepackage[style=phys,biblabel=brackets,eprint=true,sorting=noney,
  sortcites=false]{biblatex}
\newbibmacro{string+doiurl}[1]{%
  \iffieldundef{doi}{%
    \iffieldundef{url}{#1}{\href{\thefield{url}}{#1}}%
  }{\href{https://doi.org/\thefield{doi}}{#1}}%
}
\DeclareFieldFormat{title}{\usebibmacro{string+doiurl}{\mkbibemph{#1}}}
\DeclareSortingTemplate{noney}{
  \sort{\citeorder}
}

\DefineBibliographyExtras{british}{}

\newcommand*{\e}{\mathrm{e}} 
\newcommand*{\E}{\mathrm{E}} 
\newcommand*{\R}{\mathbb R}

\newcommand*{\gl}{\accentset{\lambda}{g}}
\newcommand*{\El}{\accentset{\lambda}{\E}}
\newcommand*{\Al}{\accentset{\lambda}{A}}
\newcommand*{\Ml}{\accentset{\lambda}{M}}
\newcommand*{\Sl}{\accentset{\lambda}{S}}
\newcommand*{\Xl}{\accentset{\lambda}{X}}
\newcommand*{\Yl}{\accentset{\lambda}{Y}}
\newcommand*{\Bl}{\accentset{\lambda}{B}}

\newcommand*{\te}{\tilde{\e}}
\newcommand*{\tEl}{\accentset{\lambda}{\tilde{\E}}}

\newcommand*{\On}{\mathsf{O}} 

\newcommand*{\tok}[1][k]{\xrightarrow[#1\times]{\lambda\to0}}

\theoremstyle{plain}
\newtheorem{theorem}{Theorem}[section]
\newtheorem{proposition}[theorem]{Proposition}
\newtheorem{lemma}[theorem]{Lemma}

\theoremstyle{definition}
\newtheorem{notation}[theorem]{Notation}
\newtheorem{remark}[theorem]{Remark}

\AddToHook{env/proposition/begin}{\crefalias{theorem}{proposition}}
\AddToHook{env/lemma/begin}{\crefalias{theorem}{lemma}}
\AddToHook{env/notation/begin}{\crefalias{theorem}{notation}}
\AddToHook{env/remark/begin}{\crefalias{theorem}{remark}}

\AddToHook{cmd/appendix/after}{\crefalias{section}{appendix}}

\numberwithin{equation}{section}

\title{The Newtonian limit of orthonormal frames in metric theories of
  gravity}

\author[1,a]{Philip K. Schwartz}
\author[2,3,1,b]{Arian L. von Blanckenburg}
\affil[1]{Leibniz University Hannover,
  Institute of Theoretical Physics, \par
  Appelstraße 2, 30167 Hannover, Germany}
\affil[2]{Max Planck Institute for Gravitational Physics (Albert
  Einstein Institute), \par
  Callinstraße 38, 30167 Hannover, Germany}
\affil[3]{Leibniz University Hannover,
  Institute of Gravitational Physics, \par
  Callinstraße 38, 30167 Hannover, Germany}
\affil[a]{\normalfont\texttt{\href{mailto:philip.schwartz@itp.uni-hannover.de}
    {philip.schwartz@itp.uni-hannover.de}}}
\affil[b]{\normalfont\texttt{\href{mailto:arian.von.blanckenburg@aei.mpg.de}
    {arian.von.blanckenburg@aei.mpg.de}}}

\date{}

\begin{document}
\maketitle

\vspace*{-3.5\baselineskip}

\begin{abstract}
  \noindent
  We extend well-known results on the Newtonian limit of Lorentzian
  metrics to orthonormal frames.  Concretely, we prove that, given a
  one-parameter family of Lorentzian metrics that in the Newtonian
  limit converges to a Galilei structure, any family of orthonormal
  frames for these metrics converges pointwise to a Galilei frame,
  assuming that the two obvious necessary conditions are satisfied:
  the spatial frame must not rotate indefinitely as the limit is
  approached, and the frame's boost velocity with respect to some
  fixed reference observer needs to converge.
\end{abstract}

\section{Introduction}
\label{sec:intro}

Newton--Cartan gravity \cite{Cartan:1923_24,Friedrichs:1928,
  Trautman:1963,Trautman:1965,Dombrowski.Horneffer:1964,Kuenzle:1972,
  Kuenzle:1976,Ehlers:1981a,Ehlers:1981b,Malament:1986},
\cite[chapter~4]{Malament:2012}, \cite{Schwartz:NC_gravity}---a
differential-geometric reformulation of Newtonian gravity, exposing
its similarities and allowing an elaboration on its relation to
General Relativity (\acronym{GR})---has seen a surge of interest over
the past decade~\cite{Hartong.EtAl:2023}.  Most of this renewed
interest in Newton--Cartan gravity has been due to applications in
condensed matter physics \cite{Son:2013,Geracie.Son.EtAl:2015,
  Geracie.EtAl:2015a,Geracie.EtAl:2015b}, applications in
`non-relativistic' large-speed-of-light limits of string theory and
elsewhere in quantum gravity \cite{Andringa.EtAl:2011,
  Andringa.EtAl:2012,Christensen.EtAl:2014a,Christensen.EtAl:2014b,
  Hartong.EtAl:2015,Bergshoeff.EtAl:2015,Bergshoeff.EtAl:2023}, or its
duality to so-called Carollian physics (i.e.\ small-speed-of-light
physics, or physics on null surfaces in Lorentzian spacetimes)
\cite{Duval.EtAl:2014,Bergshoeff.EtAl:2014,Bergshoeff.EtAl:2017}.
However, most recently it has come back to its roots, namely the
geometric understanding of the Newtonian limit of \acronym{GR}: based
on Newton--Cartan gravity and extending earlier post-Newton--Cartan
approaches \cite{Dautcourt:1997,Tichy.Flanagan:2011}, a systematic
full description of the post-Newtonian expansion of \acronym{GR} in a
geometric, coordinate-free language has been developed
\cite{Van_den_Bleeken:2017,Hansen.EtAl:2019,Hansen.EtAl:2020,
  Hartong.EtAl:2023}.

Such a coordinate-free understanding of the post-Newtonian behaviour
is of fundamental conceptual interest: the conventional approach of
dealing with the Newtonian limit and post-Newtonian expansion in
concrete coordinate systems---while being, of course, well-suited for
and enormously successful in predictions in an observational context
\cite{Will:2018,Poisson.Will:2014}---simply ignores the inherently
geometric nature of gravity.  Only a formulation of such limits in
coordinate-free, geometric language can count as proper understanding.
Of course, this holds true not only for standard \acronym{GR}, but
also for modified theories of gravity that keep the geometric
character of \acronym{GR} \cite{CANTATA:2021}.  Research in this
direction, however, is still in its infancy: the only works dealing
with concrete modified theories have recently established the
Newton--Cartan-like geometric description of the Newtonian limit for
two reformulations of \acronym{GR} in modified geometric frameworks,
namely the (metric) \emph{teleparallel equivalent of \acronym{GR}}
(\acronym{TEGR}) \cite{Read.Teh:2018,Schwartz:2023} and the
\emph{symmetric teleparallel equivalent of \acronym{GR}}
(\acronym{STEGR}) \cite{Wolf.EtAl:2024}.  Complementing this concrete
approach, in order to chart the possible limiting geometries of
general metric-affine Lorentzian theories of gravity, general affine
connections in Newton--Cartan / Galilei geometry have recently been
classified \cite{Schwartz:2025}.

In several situations in gravitational pyhsics---particularly in the
formulation of teleparallel theories of gravity, but also in more
general contexts---it is convenient or even necessary to work with the
geometry's `metric' part not only in terms of the spacetime metric
itself, but also in terms of local orthonormal frames (also called
`vielbeine' or, in the case of 4 spacetime dimensions, `vierbeine' or
`tetrads').  While results on the convergence of Lorentzian to Galilei
geometry in the Newtonian limit in terms of the metric are well-known
\cite{Kuenzle:1976,Ehlers:1981a,Ehlers:1981b,Malament:1986}, so far no
corresponding results have been established for orthonormal frames.
Therefore, works discussing the Newtonian limit and post-Newtonian
expansions of metric theories of gravity in terms of orthonormal
frames \cite{Hansen.EtAl:2020,Schwartz:2023} have up to now needed to
\emph{assume} the frames to have a suitable limiting behaviour,
motivated from the behaviour of the metric.  In at least one case that
we (the authors) know of, this seems to have caused some confusion,
see our critical discussion \cite{Schwartz.vonBlanck:2024a} of the
analysis of Newtonian limits of teleparallel theories in reference
\cite{Meskhidze:2024}.

In the present paper, we close this gap: given a one-parameter family
of Lorentzian metrics that in the Newtonian limit converges to a
Galilei structure (the metric structure of Newton--Cartan gravity) in
the usual sense, and any family of corresponding Lorentzian
orthonormal frames whose velocity with respect to a fixed reference
observer converges, we prove that, up to a potential spatial rotation
depending on the limit parameter, the family of orthonormal frames
converges pointwise to a Galilei frame (appropriately rescaling the
frame fields with powers of the limit parameter, i.e.\ of the
causality constant / speed of light).  Put differently: we show that
up to the obvious caveats---the frame must not spatially rotate
`faster and faster' as the limit is approached, and its boost velocity
needs to converge---frames adapted to the Lorentzian metric structure
are guaranteed to converge pointwise to their Newtonian counterparts
in the Newtonian limit.

In establishing our result, we strive for as much generality as
possible, and therefore try to keep our assumptions as weak as
possible.  In particular, we aim for regularity assumptions on the
Newtonian limit that are as weak as possible.

The structure of this paper is as follows.  First, we quickly
introduce general notation and conventions in \cref{subsec:notation}.
In \cref{sec:prelim}, we discuss some general aspects of
our low-regularity convergence assumptions.  Finally, in
\cref{sec:limits} we establish and discuss our main result on the
convergence of orthonormal frames.

\subsection{Notation and conventions}
\label{subsec:notation}

Our signature convention for Lorentzian metrics is mostly plus, i.e.\
$(-+\dots+)$.  We will take the dimension of spacetime to be $n+1$,
with $n \ge 1$.

Even though the motivation for our investigations is the application
to metric theories of gravity, and therefore to spacetime manifolds
with geometric structures defined on them in terms of tensor fields,
it is sufficient to work pointwise.  That is, instead of working with
Lorentzian metrics on \emph{manifolds} and orthonormal \emph{frames}
of vector fields, we need only work on one fixed real \emph{vector
  space} of dimension $n+1$, and consider Lorentzian metrics on it
(i.e.\ Lorentzian-signature symmetric bilinear forms) and
corresponding orthonormal \emph{bases}.  The application to Lorentzian
geometry / gravity then follows by taking for the vector space the
tangent spaces $T_pM$ of the spacetime manifold $M$, and taking
frames, metrics etc.\ as tensor fields on $M$.

For denoting the components of tensors in an arbitrary unspecified
basis, we will use lowercase Greek indices.  (These may also be
understood as `abstract' indices, however when referring to tensors
themselves we will not write the indices.)

As `frame indices' labelling the elements of a concrete basis we will
use uppercase Latin letters.  In the case of an orthonormal basis for
a Lorentzian metric, we will decompose frame indices according to $(A)
= (0,a)$, using $0$ as the timelike index and lowercase Latin letters
as spatial indices running from $1$ to $n$.  For example, the
condition that a basis $(\E_A) = (\E_0, \E_a)$ be orthonormal with
respect to a Lorentzian metric $g$ reads $g(\E_A, \E_B) = \eta_{AB}$,
where $\eta_{AB}$ are the components of the Minkowski metric in
Lorentzian coordinates, i.e.\ $(\eta_{AB}) = \mathrm{diag}
(-1,1,\dots,1)$.  If the basis is adapted to a Galilei structure (see
below), we will use $t$ instead of $0$ as the timelike index.

A \emph{Galilei structure} on an $(n+1)$-dimensional real vector space
$V$ is given by a non-vanishing \emph{clock form} $\tau \in V^*$ and a
symmetric \emph{space metric} $h \in V \otimes V$ that is positive
semidefinite of rank $n$, satisfying $\tau_\mu h^{\mu\nu} = 0$, i.e.\
such that the degenerate direction of $h$ is spanned by $\tau$.  A
\emph{Galilei basis} for $(V,\tau,h)$ is a basis $(\e_A) = (\e_t,
\e_a)$ of $V$ satisfying
\begin{equation}
  \tau(\e_t) = 1 \; , \quad
  h = \delta^{ab} \e_a \otimes \e_b \; .
\end{equation}
The dual basis of $V^*$ is then of the form $(\e^A) = (\e^t, \e^a) =
(\tau, \e^a)$.  A change of Galilei basis of the form
\begin{subequations}
\begin{equation}
  \e_t \mapsto \e_t - B^a \e_a \;, \quad
  \e_a \mapsto \e_a
\end{equation}
with $B = (B^a) \in \R^n$ is called a \emph{Milne boost} or
\emph{(local) Galilei boost}\footnote{The name `\emph{local} Galilei
  boost' is commonly used for local Galilei \emph{frames} on Galilei
  manifolds, where $B$ is an $\R^n$-valued function.}  with boost
velocity $B^a \e_a$.  The corresponding change of the dual basis reads
\begin{equation}
  \e^t = \tau \mapsto \tau \;, \quad
  \e^a \mapsto \e^a + B^a \tau \; .
\end{equation}
\end{subequations}

\section{Technical preliminaries}
\label{sec:prelim}

In this section, we will fix notation for and discuss some general
properties of the convergence of one-parameter families of elements of
finite-dimensional vector spaces.

\begin{notation}
  We will mostly be concerned with one-parameter families of elements
  of some finite-dimensional real vector space (or some specific
  subset of a vector space) depending on a parameter $\lambda>0$.  We
  will denote the dependence by writing $\lambda$ as a superscript:
  the data of `a one-parameter family $\Xl$ of elements of $V$' is a
  map $(0,\varepsilon) \ni \lambda \mapsto \Xl \in V$ for some
  $\varepsilon > 0$.

  For such one-parameter families, we will be interested in the limit
  as $\lambda \to 0$.  We will not only use the standard notation $\Xl
  \xrightarrow{\lambda\to0} x$ for the existence of such a limit, but
  extend this notation as follows.  Given a one-parameter family $\Xl$
  of elements of a subset $U \subset V$ of a vector space $V$, we will
  say that the convergence of the family to $x \in V$ is \emph{of
    order $k \in \mathbb N_0$}, written
  \begin{subequations}
    \begin{equation}
      \Xl \tok x,
    \end{equation}
    if there are $x^{(1)}, \dots, x^{(k)} \in V$ such that
    \begin{equation}
      \Xl = x + \sum_{l=1}^k \lambda^l x^{(l)} + o(|\lambda|^k),
    \end{equation}
  \end{subequations}
  i.e.\ such that $\Xl - x - \sum_{l=1}^k \lambda^l x^{(l)}$ converges
  to $0$ as $\lambda\to0$ faster than $|\lambda|^k$.  Put differently,
  convergence of order $k$ means that the family have a `Taylor
  expansion' in $\lambda$ at $\lambda = 0$ to order $k$.  Note that
  `convergence of order $0$' is just convergence.
\end{notation}

\begin{remark}
  Of course, a one-parameter family $\Xl$ in $V$ converging (of order
  $0$) as $\lambda \to 0$ means that the map $(0,\varepsilon) \ni
  \lambda \mapsto \Xl$ may be extended to a map on the half-open
  interval $[0,\varepsilon)$ that is continuous at $0$.  For
  convergence of order $k > 0$, by Taylor's theorem it is sufficient
  that this map be $k$ times differentiable at $0$.  However, for $k
  \ge 2$, this is not necessary, i.e.\ order-$k$ convergence (the
  existence of an order-$k$ Taylor expansion at $0$) is weaker than
  $k$-fold differentiability at $0$: it may be the case that none of
  the derivatives higher than the first actually exist.\footnote{For
    example \cite{StackExchange:2020}, consider the function $f\colon
    [0,\infty) \to \R$ defined by
    \begin{subequations}
      \begin{align}
        f(\lambda) &:= \begin{cases}
          0 & \lambda = 0,\\
          \lambda^{k+1} \sin(\lambda^{-k}) & \lambda > 0.
        \end{cases}\\
        \shortintertext{It satisfies $f(\lambda) = o(|\lambda|^k)$
        (since $\lim_{\lambda\to0} \lambda \sin(\lambda^{-k}) = 0$),
        i.e.\ $f(\lambda) \tok 0$.  But its first derivative is}
        f'(\lambda) &= \begin{cases}
          0 & \lambda = 0,\\
          (k+1) \lambda^k \sin(\lambda^{-k}) - k \cos(\lambda^{-k}) &
          \lambda > 0,
        \end{cases}
      \end{align}
    \end{subequations}
    which is not continuous at $\lambda = 0$, such that $f''(0)$
    cannot exist.}
\end{remark}

\begin{remark}
  Let $\Xl$ be a one-parameter family of elements of a
  finite-dimensional real vector space $V$, satisfying $\Xl \tok x$.
  Further, let $U \subset V$ be an open neighbourhood of $x$ and let
  $f \colon U \to W$ be a map to a finite-dimensional real vector
  space $W$ that has a Taylor expansion at $x$ to order $k$, i.e.\
  such that there are multilinear maps $F^{(l)} \colon V^l \to W$, $l
  = 1, \dots, k$, satisfying
  \begin{equation}
    f(x + v) = f(x) + \sum_{l=1}^k
    F^{(l)}(\underbrace{v,\dots,v}_{\text{$l$ times}}) + o(\|v\|^k)
  \end{equation}
  with respect to any norm on $V$.  (Extending our previously
  introduced notation, we might write this assumption as $f(x + v)
  \xrightarrow[k\times]{\|v\| \to 0} f(x)$.)  Then, by inserting the
  Taylor expansion for $\Xl$ at $\lambda = 0$ into this expansion, we
  directly obtain that $f(\Xl) \tok f(x)$.

  In particular, if $\Xl \tok x$ and $f$ is $k$ times differentiable
  (or even $C^\infty$) at $x$, we have $f(\Xl) \tok f(x)$---i.e.\
  \emph{$k$-fold differentiable maps preserve order-$k$ convergence}.
\end{remark}

In the arguments in the remainder of the paper, we will regularly use
this observation.  First, we will use it to prove a `square root'
lemma on one-parameter families of matrices:
\begin{lemma}
  \label{lemma:limit_matrix}
  Let $\Ml$ be a one-parameter family of real-valued symmetric
  $n\times n$ matrices satisfying $\Ml \tok \mathbb 1$, where $\mathbb
  1$ is the identity matrix.  Then for $\lambda$ small enough there is
  a one-parameter family $\Sl$ of symmetric matrices satisfying $\Sl
  \tok \mathbb 1$ and $\Ml = \Sl^2$.

  \begin{proof}
    The matrix exponential $\exp \colon \mathfrak{gl}(n) \to
    \mathsf{GL}(n)$ is a diffeomorphism from a sufficiently small open
    neighbourhood $U \subset \mathfrak{gl}(n)$ of the zero matrix onto
    its image $\exp(U) \subset \mathsf{GL}(n)$, which is an open
    neighbourhood of the identity matrix $\mathbb 1$.  Its inverse we
    denote by $\log \colon \exp(U) \to U$.

    As one easily sees by a diagonalisation argument, any positive
    definite symmetric matrix has \emph{a} logarithm that is
    symmetric.  For a positive definite symmetric matrix $X$, the
    eigenvalues of its symmetric logarithm are the logarithms of the
    eigenvalues of $X$; in particular, for a symmetric matrix $X$
    close enough to $\mathbb 1$, its symmetric logarithm will lie in
    $U$.  Hence, we may assume (by perhaps shrinking $U$) that for all
    $X \in \exp(U)$ that are symmetric, $\log(X) \in U$ is also
    symmetric.

    Since $\Ml$ converges to $\mathbb 1$, for $\lambda$ sufficiently
    small we have $\Ml \in \exp(U)$, such that we can define
    $\accentset{\lambda}{m} := \log(\Ml)$.  We now define $\Sl :=
    \exp(\frac{1}{2} \accentset{\lambda}{m})$, such that by the
    Baker--Campbell--Hausdorff formula we have $\Sl^2 =
    \exp(\accentset{\lambda}{m}) = \Ml$.  Since the $\Ml$ are
    symmetric, by our assumption on $U$ the $\accentset{\lambda}{m}$
    are symmetric as well.  Expressing the exponential as a power
    series then shows that the $\Sl$ are symmetric.  Since $\Ml \tok
    \mathbb 1$ and $\log$ is $C^\infty$ (on all of $\exp(U)$, and in
    particular at $\mathbb 1$), we have $\accentset{\lambda}{m} \tok
    0$; since $\exp$ is $C^\infty$, this finally shows $\Sl \tok
    \mathbb 1$.
  \end{proof}
\end{lemma}

We will also need a `square root' lemma for tensor products of
one-parameter families of vectors:
\begin{lemma}
  \label{lemma:limit_tens_prod}
  Let $\Xl$ be a one-parameter family of elements of a
  finite-dimensional real vector space $V$.  Assume that the limit
  $\accentset{0}{X} := \lim_{\lambda\to0} \Xl$ exists and is
  non-vanishing.  Writing $\Bl = \Xl \otimes \Xl$, assume that $\Bl
  \tok \accentset{0}{B}$.  Then $\Xl \tok \accentset{0}{X}$ as well.

  \begin{proof}
    We write $\e_1 := \accentset{0}{X}$ and extend this to a basis
    $(\e_1, \dots, \e_n)$ of $V$.  Expressed in terms of this basis,
    we then have $\accentset{0}{B} = \e_1 \otimes \e_1$.

    We expand $\Xl$ and $\Bl$ as $\Xl = \Xl^a \e_a$, $\Bl = \Bl^{ab}
    \e_a \otimes \e_b$, such that we have $\Bl^{ab} = \Xl^a \Xl^b$ for
    all $a,b$.  We also know that
    \begin{equation}
      \lim_{\lambda\to0} \Bl^{ab}
      = \begin{cases}
        1 & a = b = 1, \\
        0 & \text{otherwise}.
      \end{cases}
    \end{equation}

    In particular, we have $(\Xl^1)^2 = \Bl^{11}$.  This implies that
    $\Xl^1 = \pm \sqrt{\Bl^{11}}$ with a potentially
    $\lambda$-dependent sign.  Since the limit $\lim_{\lambda\to0}
    \Xl^1 = 1$ exists and is positive, for $\lambda$ small enough we
    have $\Xl^1 = \sqrt{\Bl^{11}}$.  Since $\Bl^{11} \tok
    \accentset{0}{B}^{11} = 1 \ne 0$ and the square root function is
    $C^\infty$ at $1$, this implies that $\Xl^1 \tok 1 =
    \accentset{0}{X}^1$.

    Since the inversion function $x \mapsto 1/x$ is $C^\infty$ at $1$,
    this further implies $1/\Xl^1 \tok 1$.  Combining this with the
    fact that for $a > 1$ we have $\Xl^a = \Bl^{1a} / \Xl^1$, we
    finally obtain $\Xl^a \tok \accentset{0}{X}^a$ as well.
  \end{proof}
\end{lemma}

\pagebreak

\section{Limits of orthonormal frames}
\label{sec:limits}

In this section we are going to discuss the Newtonian limit of
orthonormal frames.  As explained in the introduction, for our
discussion we work on an $(n+1)$-dimensional real vector space $V$,
such that in fact we will consider the limit of orthonormal
\emph{bases}; to apply our results to theories of gravity, one needs
to take $V = T_pM$ the tangent spaces of the spacetime manifold $M$,
and take frames, metrics etc.\ as tensor fields on $M$.  Note however
that this means that our results, when applied to the manifold case,
allow no conclusion on the smoothness of the limiting frame---any such
conclusion needs extra assumptions on the convergence of derivatives.

First we recall a standard result on the Newtonian limit from
Lorentzian metrics to Galilei structures \cite{Kuenzle:1976,
  Ehlers:1981a,Ehlers:1981b,Malament:1986}, which we present in a
quite general formulation:
\begin{proposition}
  \label{prop:metr_conv_Galilei}
  Let $\gl$ be a one-parameter family of Lorentzian metrics on an
  $(n+1)$-dimensional real vector space $V$ that satisfies
  \begin{enumerate}[label=(\roman*)]
  \item \label{item:conv_metr} $\lambda\gl \tok -\tau \otimes \tau$
    for some $\tau \in V^*$, and
  \item \label{item:conv_inv_metr} $\gl^{-1} \tok h$ for some
    (symmetric) $h \in V \otimes V$,
  \end{enumerate}
  for some $k \in \mathbb N_0$.  (As limit of a symmetric bilinear
  form, $h$ is automatically symmetric.)  Then if any of the following
  conditions holds, $\tau$ and $h$ define a Galilei structure on $V$:
  \begin{enumerate}[label=(\alph*)]
  \item \label{item:metr_conv_assump_h} $k \ge 1$ and $h$ has rank $n$.
  \item \label{item:metr_conv_assump_tau} $k \ge 1$ and $\tau$ is
    non-vanishing.
  \item \label{item:metr_conv_assump_both} $\tau$ is non-vanishing,
    and $h$ has rank $n$ and is positive semidefinite.
  \end{enumerate}
\end{proposition}
Before proving this result, we want to remark on its specific
formulation that we decided to give here.  The parameter $\lambda$
parametrising the Newtonian limit as it tends to zero is to be
interpreted as the \emph{causality constant} of the spacetime, with $c
= \frac{1}{\sqrt{\lambda}}$ being the speed of light.  In the
literature \cite{Kuenzle:1976,Ehlers:1981a,Ehlers:1981b,
  Malament:1986}, it is common to assume the limits
\ref{item:conv_metr}, \ref{item:conv_inv_metr} to be differentiable in
$\lambda$ at $\lambda = 0$ (i.e.\ $k \ge 1$) and assume the signature
of $h$, i.e.\ consider condition \ref{item:metr_conv_assump_h} with
the additional assumption that $h$ be positive semidefinite.  We
already weaken this by not requiring the semidefiniteness.  Instead,
condition \ref{item:metr_conv_assump_tau} yields the same conclusion,
replacing the assumption on $h$ by that of non-vanishing $\tau$.  This
is interesting for two reasons: on the one hand, the assumption on
$\tau$ might a priori seem weaker that that on $h$; on the other hand,
physically speaking it is an assumption on the limit of temporal
durations instead of spatial lengths.  To our knowledge, this
alternative formulation of the limit assumption has not appeared in
the literature before.  Finally, condition
\ref{item:metr_conv_assump_both} allows for $k = 0$, i.e.\ assumes
only \emph{existence} of the limits in \ref{item:conv_metr},
\ref{item:conv_inv_metr}, thus reducing the regularity assumption on
the limiting behaviour as much as possible, at the expense of needing
to assume the signatures of both $\tau\otimes\tau$ and $h$.

\begin{proof}[Proof of \cref{prop:metr_conv_Galilei}]
  For all three cases we have to show that $\tau_\mu h^{\mu\nu} = 0$;
  in case \ref{item:metr_conv_assump_h} we additionally have to show
  that $\tau$ is non-vanishing, in case
  \ref{item:metr_conv_assump_tau} that $h$ has rank $n$, and in both
  these cases that $h$ is positive semidefinite.

  First note that \ref{item:conv_metr} and \ref{item:conv_inv_metr}
  for any $k$ imply that $\lambda \gl_{\mu\nu} = -\tau_\mu \tau_\nu +
  o(\lambda^0)$ and $\gl^{\mu\nu} = h^{\mu\nu} + o(\lambda^0)$.  By
  definition of the inverse metric, this yields $\lambda
  \delta^\mu_\rho = \gl^{\mu\nu} \lambda \gl_{\nu\rho} = - h^{\mu\nu}
  \tau_\nu \tau_\rho + o(\lambda^0)$, implying $h^{\mu\nu} \tau_\nu
  \tau_\rho = 0$.  In cases \ref{item:metr_conv_assump_both} and
  \ref{item:metr_conv_assump_tau}, $\tau$ is non-vanishing, so we
  obtain $\tau_\mu h^{\mu\nu} = 0$.

  This finishes the proof of case \ref{item:metr_conv_assump_both}.
  For the proof of the remaining cases \ref{item:metr_conv_assump_h}
  and \ref{item:metr_conv_assump_tau} we now assume that $k \ge 1$.
  Then \ref{item:conv_metr} and \ref{item:conv_inv_metr} imply that
  $\lambda \gl_{\mu\nu} = -\tau_\mu \tau_\nu + \lambda
  g^{(1)}_{\mu\nu} + o(|\lambda|)$ and $\gl^{\mu\nu} = h^{\mu\nu} +
  \lambda m^{\mu\nu} + o(|\lambda|)$.  Again by definition of the
  inverse metric, we have
  \begin{align}
    \lambda \delta^\mu_\rho
    &= \gl^{\mu\nu} \lambda \gl_{\nu\rho} \nonumber\\
    &= \left(h^{\mu\nu} + \lambda m^{\mu\nu} + o(|\lambda|)\right)
      \left(-\tau_\nu \tau_\rho + \lambda g^{(1)}_{\nu\rho}
        + o(|\lambda|)\right) \nonumber\\
    &= \lambda \left(h^{\mu\nu} g^{(1)}_{\nu\rho} - m^{\mu\nu}
      \tau_\nu \tau_\rho \right) + o(|\lambda|),\\
    \intertext{where we used that $h^{\mu\nu} \tau_\nu \tau_\rho = 0$.
    Comparison of coefficients now implies}
    \label{eq:metr_conv_delta}
    \delta^\mu_\rho
    &= h^{\mu\nu} g^{(1)}_{\nu\rho} - m^{\mu\nu} \tau_\nu \tau_\rho \;
      .
  \end{align}
  The left-hand side of \eqref{eq:metr_conv_delta} has full rank
  (namely $n + 1$).  We may now use this to prove the remaining
  statements about the ranks of $\tau$ and $h$ in the cases
  \ref{item:metr_conv_assump_h} and \ref{item:metr_conv_assump_tau}:
  \begin{enumerate}[label=(\alph*)]
  \item Since $h$ has rank $n$, the first term on the right-hand side
    of \eqref{eq:metr_conv_delta} has rank at most $n$.  Thus the
    second needs to have rank at least $1$.  This implies that $\tau$
    has rank $1$, i.e.\ is non-vanishing.  Using this, $h^{\mu\nu}
    \tau_\nu \tau_\rho = 0$ again implies $\tau_\mu h^{\mu\nu} = 0$.
  \item Since we already know that $\tau_\mu h^{\mu\nu} = 0$, we know
    that $h$ has rank at most $n$.  Hence the second term on the
    right-hand side of \eqref{eq:metr_conv_delta} needs to have rank
    at least $1$ (otherwise the sum could not have rank $n+1$); since
    it is proportional to $\tau_\rho$, it has rank $1$.  Hence the
    first term needs to have rank at least $n$, showing that $h$ has
    rank $n$.
  \end{enumerate}

  In both cases, it remains to prove that $h$ is positive
  semidefinite, i.e.\ that it is positive definite in its $n$
  non-degenerate directions.  First we observe that contracting
  \eqref{eq:metr_conv_delta} with $\tau_\mu$, we obtain $-1 = \tau_\mu
  m^{\mu\nu} \tau_\nu$.  This shows that $\gl^{-1} (\tau,\tau) = -
  \lambda + o(|\lambda|)$, such that for $\lambda>0$ sufficiently
  small $\tau$ is timelike in the Lorentzian sense with respect to
  $\gl$.

  Now consider a $\beta \in V^*$ satisfying $h(\beta,\beta) \ne 0$.
  Since $h(\tau,\tau) = 0$, we know that $\beta$ and $\tau$ are
  linearly independent; hence the projection $\tilde{\beta} := \beta -
  \frac{\gl^{-1}(\tau,\beta)}{\gl^{-1}(\tau,\tau)} \tau$ of $\beta$
  onto the $\gl^{-1}$-orthogonal complement of $\tau$ is non-zero.
  Since for $\lambda$ small enough $\tau$ is timelike, $\tilde{\beta}$
  is spacelike (both in the Lorentzian sense).  A direct computation
  shows that $\tilde{\beta} = \beta + m(\tau,\beta) \tau +
  o(\lambda^0)$.  Since $\tilde{\beta}$ is spacelike for $\lambda$
  small enough, this implies $0 < \gl^{-1}(\tilde{\beta},
  \tilde{\beta}) = h(\beta,\beta) + o(\lambda^0)$.  Hence we have
  $h(\beta, \beta) > 0$, showing that $h$ is indeed positive definite
  in its non-degenerate directions.
\end{proof}

Now we turn to our main result, which concerns the Newtonian limit of
Lorentzian orthonormal bases.  Again we aim for a formulation with
assumptions that are as weak as possible.

\begin{theorem}
  \label{theorem:conv_frame}
  Let $\gl$ be a one-parameter family of Lorentzian metrics on an
  $(n+1)$-dimensional real vector space $V$ that satisfies
  \begin{enumerate}[label=(\roman*)]
  \item \label{item:conv_metr_frame} $\lambda\gl \tok[0] -\tau \otimes
    \tau$ for a non-vanishing $\tau \in V^*$, and
  \item \label{item:conv_inv_metr_frame} $\gl^{-1} \tok h$ for a rank
    $n$ positive-semidefinite symmetric $h \in V \otimes V$,
  \end{enumerate}
  for some $k \in \mathbb N_0$.  Let $(\e_A)$ be a Galilei basis for
  $(V,\tau,h)$.  Further, let $(\El_A)$ be a one-parameter family of
  Lorentzian orthonormal bases for the metrics $\gl$, i.e.\ such that
  for each value of $\lambda > 0$ we have $\gl(\El_A, \El_B) =
  \eta_{AB}$, and assume that the limit $\lim_{\lambda\to0}
  \frac{1}{\sqrt{\lambda}} \El_0$ exists and we have
  $\frac{1}{\sqrt{\lambda}} \El_0 \tok[l] \lim_{\lambda\to0}
  \frac{1}{\sqrt{\lambda}} \El_0$ for some $l \in \mathbb N_0$.

  Then there is a one-parameter family $\Al =
  (\tensor{\Al}{^a_b})_{a,b = 1}^n$ of matrices in $\On(n)$ and a
  vector $B = (B^a) \in \R^n$ such that
  \begin{subequations}
  \begin{align}
    \label{eq:conv_dual_frame}
    \sqrt{\lambda} \El^0
    &\tok[\min(k,l)] \pm \e^t = \pm \tau \; ,
    &\tensor{(\Al^{-1})}{^a_b} \El^b
    &\tok[\min(k,l)] \e^a + B^a \tau \; ,\\
    \label{eq:conv_frame}
    \frac{1}{\sqrt{\lambda}} \El_0
    &\tok[l] \pm (\e_t - B^a \e_a) \; ,
    &\tensor{\Al}{^a_b} \El_a
    &\tok[\min(k,l+1)] \e_b \; .
  \end{align}
  \end{subequations}

  Put differently, up to a sign change of $\El_0$, a rotation of the
  spacelike basis $(\El_a)$, and a Milne boost, the Lorentzian basis
  and dual basis, properly rescaled by powers of $\lambda$, converge
  to the Galilei basis and dual basis.
\end{theorem}

Before proving this result, we are going to give a few remarks on its
convergence assumptions, in particular regarding the assumed orders of
convergence.

\begin{remark}
  \begin{enumerate}[label=(\alph*)]
  \item The assumption that $\frac{1}{\sqrt{\lambda}} \El_0$ converges
    as $\lambda \to 0$ means that the velocity parameter of the
    Lorentz boost linking the two Lorentzian states of motion $w :=
    \e_t/\sqrt{-\gl(\e_t,\e_t)}$ and $\pm\El_0$ converges.  More
    explicitly, this may be seen as follows.

    By assumption \ref{item:conv_metr_frame}, for $\lambda$
    sufficiently small the timelike basis vector $\e_t$ of the Galilei
    basis is timelike in the Lorentzian sense with respect to $\gl$;
    i.e.\ it represents a Lorentzian observer's state of motion.
    Normalising $\e_t$, we obtain the Lorentzian unit timelike vector
    $w$ as above.  By construction, $\frac{1}{\sqrt{\lambda}} w$
    converges to $\e_t$.

    Now given the two unit timelike vectors $w$ and $\pm\El_0$, where
    we choose the sign such that $w$ and $\pm\El_0$ point in the same
    time direction (i.e.\ $\gl(w,\pm\El_0) < 0$), there is a unique
    Lorentz transformation of $(V,\gl)$ that (1) maps $w$ to
    $\pm\El_0$, and (2) is a boost with respect to $w$, i.e.\ acts
    trivially on a spacelike $(n-1)$-plane orthogonal to
    $w$.\footnote{Instead of (2), we might of course also characterise
      this transformation by it being a boost with respect to
      $\pm\El_0$.}  The boost velocity vector in
    $\mathrm{span}\{w\}^\perp$ characterising this boost is given by
    \begin{subequations}
    \label{eq:boost_vel}
    \begin{align}
      v_\text{boost}
      &= \frac{\frac{1}{\sqrt{\lambda}} \El_0}{-\lambda
          \gl\Big( \frac{1}{\sqrt{\lambda}} w,
            \frac{1}{\sqrt{\lambda}} \El_0 \Big)}
        - \frac{1}{\sqrt{\lambda}} w \; ; \\
      \shortintertext{conversely, we have}
      \frac{1}{\sqrt{\lambda}} \El_0
      &= \pm \frac{\frac{1}{\sqrt{\lambda}} w + v_\text{boost}}{\sqrt{1
        - \lambda \gl(v_\text{boost},v_\text{boost})}} \; .
    \end{align}
    \end{subequations}
    (This uses that $\frac{1}{\sqrt{\lambda}}$ is interpreted as the
    speed of light; for details, see
    \cref{sec:appendix_boost_velocity}.)  This shows that
    $\frac{1}{\sqrt{\lambda}} \El_0$ converges as $\lambda \to 0$ if
    and only if $v_\text{boost}$ converges, thus providing a clear
    physical motivation for the assumption that
    $\frac{1}{\sqrt{\lambda}} \El_0$ converges.

    Using \eqref{eq:conv_frame}, a direct calculation further shows
    that $v_\text{boost} \xrightarrow{\lambda\to0} -B^a \e_a$: the
    Lorentzian boost velocity between $w$ and $\pm \El_0$ converges to
    the Milne boost velocity between $\e_t$ and the limit of
    $\frac{1}{\sqrt{\lambda}} \El_0$.
  \item Even though in \cref{theorem:conv_frame} we only assume that
    the rescaled metric $\lambda\gl$ converge to $-\tau\otimes\tau$ at
    all (i.e.\ we assume convergence of order $0$), it actually
    \emph{follows} that this convergence is of order $\min(k,l)$: we
    can express the rescaled metric as $\lambda\gl = \lambda \eta_{AB}
    \El^A \otimes \El^B = - \sqrt{\lambda} \El^0 \otimes
    \sqrt{\lambda} \El^0 + \lambda \delta_{ab}
    \tensor{(\Al^{-1})}{^a_c} \El^c \otimes \tensor{(\Al^{-1})}{^b_d}
    \El^d$, such that \eqref{eq:conv_dual_frame} implies convergence
    of order $\min(k,l)$.
  \item If we know a priori that the convergence of $\lambda\gl$ to
    $-\tau\otimes\tau$ is of order $m > k$, we can use this to improve
    on the order of convergence for the timelike dual basis vector as
    stated in \eqref{eq:conv_dual_frame}: we can express the timelike
    dual basis vector in terms of the metric and the timelike basis
    vector as $\sqrt{\lambda} \El^0 = -
    \lambda\gl(\frac{1}{\sqrt{\lambda}} \El_0, \cdot)$.  Together with
    the assumed order-$l$ convergence of $\frac{1}{\sqrt{\lambda}}
    \El_0$, this shows that $\sqrt{\lambda} \El^0$ converges of order
    $\min(m,l)$, improving on the order $\min(k,l)$ from
    \eqref{eq:conv_dual_frame} if $m > k$ and $k < l$.\footnote{Note
      that for the \emph{spacelike} dual basis vectors, the
      corresponding argument would not improve our knowledge of the
      order of convergence: according to \eqref{eq:conv_frame} the
      convergence order of the spacelike basis vectors is already
      bounded above by $k$, such that we cannot `get rid of' this $k$
      dependence and thus improve the order.}
  \item In addition to the previous argument, we can \emph{also} use a
    priori knowledge of order-$m$ convergence of $\lambda\gl$ to
    improve on the order of convergence of $\sqrt{\lambda} \El^0$ in a
    different way: we can write $\sqrt{\lambda} \El^0 \otimes
    \sqrt{\lambda} \El^0 = - \lambda\gl + \lambda \delta_{ab}
    \tensor{(\Al^{-1})}{^a_c} \El^c \otimes \tensor{(\Al^{-1})}{^b_d}
    \El^d$.  By \eqref{eq:conv_dual_frame}, the second term on the
    right-hand side of this equation converges of order $\min(k,l) +
    1$.  Combining this with the assumption that $\lambda\gl$ converge
    of order $m$, we obtain that $\sqrt{\lambda} \El^0 \otimes
    \sqrt{\lambda} \El^0$ converges to $\tau \otimes \tau$ of order
    $\min(m, \min(k,l) + 1)$.  Hence, we may apply
    \cref{lemma:limit_tens_prod}, which shows that $\sqrt{\lambda}
    \El^0$ converges of order $\min(m, \min(k,l) + 1)$.

    Note that it depends on the concrete values of $k, l, m$ which of
    these arguments yields the strongest result for the order of
    convergence of $\sqrt{\lambda} \El^0$ (i.e.\ which of $\min(k,l)$,
    $\min(m,l)$, $\min(m, \min(k,l) + 1)$ is largest).
  \end{enumerate}
\end{remark}

\begin{proof}[Proof of \cref{theorem:conv_frame}]
  We write $X := \lim_{\lambda\to0} \frac{1}{\sqrt{\lambda}} \El_0$;
  by assumption, we have ${\frac{1}{\sqrt{\lambda}} \El_0 \tok[l] X}$.
  Combining this with assumption \ref{item:conv_metr_frame}, we obtain
  $\gl(\El_0,\El_0) \tok[0] -(\tau(X))^2$.  Since $\gl(\El_0,\El_0) =
  -1$, this shows that $(\tau(X))^2 = 1$, i.e.\ the $\tau$ component
  of $X$ is $\tau(X) = \pm 1$.  Writing the `spatial' components as
  $\e^a(X) =: -\tau(X) B^a$, we thus have
  \begin{equation}
    \frac{1}{\sqrt{\lambda}} \El_0 \tok[l] \pm (\e_t - B^a \e_a) \; .
  \end{equation}

  Expressing the inverse metric in terms of the Lorentzian orthonormal
  basis, assumption \ref{item:conv_inv_metr_frame} becomes
  \begin{equation}
    -\El_0 \otimes \El_0 + \delta^{ab} \El_a \otimes \El_b \tok h \;
    .
  \end{equation}
  Our assumption on the convergence of $\frac{1}{\sqrt{\lambda}}
  \El_0$ implies that $\frac{1}{\lambda} \El_0 \otimes \El_0 \tok[l]
  (\text{something})$, such that $\El_0 \otimes \El_0 \tok[(l+1)] 0$.
  Combined, this shows
  \begin{equation}
    \label{eq:spatial_inv_metr_conv}
    \delta^{ab} \El_a \otimes \El_b \tok[\tilde{k}] h \; ,
  \end{equation}
  where $\tilde{k} := \min(k,l+1)$.  Applying
  \eqref{eq:spatial_inv_metr_conv} to $\e^c$ and $\e^d$ and writing
  $\tensor{\Yl}{^a_b} := \e^a(\El_b)$, we obtain
  \begin{subequations}
  \begin{align}
    \delta^{ab} \tensor{\Yl}{^c_a} \tensor{\Yl}{^d_b}
    &\tok[\tilde{k}] \delta^{cd} \; . \\
    \shortintertext{In matrix notation, this reads}
    \label{eq:YY^T_conv}
    \Yl \Yl^T
    &\tok[\tilde{k}] \mathbb 1 \; .
  \end{align}
  \end{subequations}
  Therefore, by \cref{lemma:limit_matrix} for $\lambda$ small enough
  there is a one-parameter family $\Sl$ of symmetric matrices
  satisfying $\Sl \tok[\tilde{k}] \mathbb 1$ and $\Yl \Yl^T = \Sl^2$.
  
  \Cref{eq:YY^T_conv} implies that for $\lambda$ small enough $\Yl$
  has full rank, i.e.\ it is invertible.  Hence, the equation for
  $\Sl^2$ is equivalent to $\Yl^{-1} \Sl (\Yl^{-1} \Sl)^T = \mathbb
  1$, i.e.\ $\Al := \Yl^{-1} \Sl \in \On(n)$.\footnote{We are only
    interested in the limiting behaviour as $\lambda \to 0$, hence it
    does not matter that the expression for $\Al$ is only defined for
    $\lambda$ small enough.  For larger $\lambda$, we may take for
    $\Al$ an arbitrary matrix in $\On(n)$.}  This also shows that for
  $\lambda$ small enough, $\Sl$ is invertible (with inverse $\Al^{-1}
  \Yl^{-1}$).  With $\Sl \tok[\tilde{k}] \mathbb 1$, we have $\Sl^{-1}
  \tok[\tilde{k}] \mathbb 1$ (since the entries of the inverse matrix
  are rational functions in the entries of the original matrix).

  Applying \eqref{eq:spatial_inv_metr_conv} to $\tau$ and $\e^c$, we
  obtain
  \begin{subequations}
  \begin{align}
    \delta^{ab} \tau(\E_a) \tensor{\Yl}{^c_b}
    &\tok[\tilde{k}] 0 \; . \\
    \shortintertext{Contracting this with $\tensor{(\Sl^{-1})}{^d_c}$
    and using $\Sl^{-1} \tok[\tilde{k}] \mathbb 1$ yields}
    \delta^{ab} \tau(\E_a) \tensor{(\Sl^{-1} \Yl)}{^d_b}
    &\tok[\tilde{k}] 0 \; . \\
    \shortintertext{Now $\Sl^{-1} \Yl = \Al^{-1}$, and $\Al \in
    \On(n)$ means $\tensor{(\Al^{-1})}{^d_b} \delta^{ab} =
    \tensor{\Al}{^a_c} \delta^{dc}$.  Thus we have}
    \tensor{\Al}{^a_c} \delta^{dc} \tau(\E_a)
    &\tok[\tilde{k}] 0 \; , \\
    \shortintertext{i.e.}
    \label{eq:rot_tau_Ea_conv_0}
    \tensor{\Al}{^a_c} \tau(\E_a)
    &\tok[\tilde{k}] 0 \; .
  \end{align}
  \end{subequations}
  Expressing the spacelike Lorentzian basis vectors $\El_a$ in terms
  of the Galilei basis $(\e_A)$, we obtain
  \begin{subequations}
  \begin{align}
    \El_a
    &= \tau(\El_a) \e_t + \e^c(\El_a) \e_c
    = \tau(\El_a) \e_t + \tensor{\Yl}{^c_a} \e_c \; . \\
    \intertext{Contracting this equation with $\tensor{\Al}{^a_b}$
    yields}
    \tensor{\Al}{^a_b} \El_a
    &= \tensor{\Al}{^a_b} \tau(\El_a) \e_t
      + \underbrace{\tensor{(\Yl \Al)}{^c_b}}_{= \tensor{\Sl}{^c_b}}
        \e_c
    = \tensor{\Al}{^a_b} \tau(\El_a) \e_t
      + \tensor{\Sl}{^c_b} \e_c \; .
  \end{align}
  Combining this with \eqref{eq:rot_tau_Ea_conv_0} and $\Sl
  \tok[\tilde{k}] \mathbb 1$ gives
  \begin{equation}
    \tensor{\Al}{^a_b} \El_a
    \tok[\tilde{k}] \e_b \; .
  \end{equation}
  \end{subequations}

  So far, we have proved \eqref{eq:conv_frame} on the convergence of
  the frame; we will now use this to prove convergence of the dual
  frame according to \eqref{eq:conv_dual_frame}.  For this, we use the
  following general observation:
  \begin{lemma}
    \label{lemma:conv_gen_basis}
    Let $(\te_A)$ be a basis of a finite-dimensional real vector space
    $V$.  Let $(\tEl_A)$ be a one-parameter family of bases of $V$,
    converging to $(\te_A)$ according to
    \begin{align}
      \label{eq:conv_gen_basis}
      \tEl_A &\tok[k_A] \te_A \\
      \shortintertext{with orders $k_A \in \mathbb{N}_0$.  Then the
      family $(\tEl^A)$ of dual bases of $V^*$ converges to the dual
      basis $(\te^A)$ with order $\hat k = \min\{k_A\}$, i.e.}
      \label{eq:conv_gen_dual_basis}
      \tEl^A &\tok[\hat k] \te^A \; .
    \end{align}

    \begin{proof}
      We denote by $\Ml$ the basis change matrix from $(\te_A)$ to
      $(\tEl_A)$, which is defined by $\tEl_A = \tensor{\Ml}{^B_A}
      \te_B$.  \Cref{eq:conv_gen_basis} then means that $\Ml$
      converges to $\mathbb 1$ as $\lambda \to 0$, the $A$-th column
      converging of order $k_A$.  Hence, the matrix converges
      according to $\Ml \tok[\hat k] \mathbb 1$, implying $\Ml^{-1}
      \tok[\hat k] \mathbb 1$.  Since the dual bases satisfy $\tEl^A =
      \tensor{(\Ml^{-1})}{^A_B} \te^B$, this shows
      \eqref{eq:conv_gen_dual_basis}.\looseness-1
    \end{proof}
  \end{lemma}
  Applying \cref{lemma:conv_gen_basis} to the bases $\tEl_0 :=
  \frac{1}{\sqrt{\lambda}} \El_0$, $\tEl_b := \tensor{\Al}{^a_b}
  \El_a$ and $\te_t := \pm (\e_t - B^a \e_a)$, $\te_a := \e_a$ shows
  convergence of the dual frame according to
  \eqref{eq:conv_dual_frame}.  This finishes the proof of
  \cref{theorem:conv_frame}.
\end{proof}

\section*{Acknowledgements}

We thank Domenico Giulini for valuable discussions.

\printbibliography

\appendix

\section{The boost velocity relating two Lorentzian states of motion}
\label{sec:appendix_boost_velocity}

\newcommand*{\B}{\mathcal B}

In the following, we are going to discuss in detail the relation
between two Lorentzian states of motion and the velocity of the boost
relating them.

Let $(V,g)$ be a finite-dimensional real vector space with a
Lorentzian metric, and let $w, \tilde{w} \in V$ be two unit timelike
vectors pointing in the same time direction, i.e.\ two vectors
satisfying
\begin{equation}
  g(w,w) = -1, \quad g(\tilde{w},\tilde{w}) = -1, \quad 
  g(w,\tilde{w}) < 0.
\end{equation}
We know that there is a unique Lorentz transformation $\B$ of $(V,g)$
that maps $w$ to $\tilde{w}$ and is a boost with respect to $w$.  This
boost is characterised by its \emph{rapidity} $\theta \in \R$ and its
unit spacelike \emph{direction} (from the point of view of $w$) $d \in
\mathrm{span}\{w\}^\perp$, $g(d,d) = 1$.  Concretely, in terms of
these the boost acts according to
\begin{subequations}
\begin{align}
  \label{eq:boost_rap_dir_1}
  \B(w) &= \cosh(\theta) w + \sinh(\theta) d = \tilde{w}, \\
  \B(d) &= \sinh(\theta) w + \cosh(\theta) d, \\
  \B(u) &= u \; \text{for} \; u \in \mathrm{span}\{w,d\}^\perp .
\end{align}
\end{subequations}
The speed of the boost, measured in units of the speed of light, is
given in terms of the rapidity as $\tanh(\theta)$.  Therefore, the
boost's spacelike velocity vector, divided by the speed of light, is
$\tanh(\theta) d$.  From \eqref{eq:boost_rap_dir_1} we directly obtain
$\cosh(\theta) = -g(w,\tilde{w})$, such that we may compute the boost
velocity divided by the speed of light in terms of $w$ and $\tilde{w}$
as
\begin{align}
  \frac{v_\B}{c}
  &= \tanh(\theta) d = \frac{\sinh(\theta)}{\cosh(\theta)} d
    \nonumber\\
  &= \frac{\tilde{w} - \cosh(\theta) w}{\cosh(\theta)}
    = \frac{\tilde{w}}{\cosh(\theta)} - w \nonumber\\
  &= \frac{\tilde{w}}{-g(w,\tilde{w})} - w .
\end{align}
Conversely, we may express the scalar product of $w$ and
$\tilde{w}$---which is nothing but the boost's `$\gamma$ factor'---in
terms of the boost velocity as
\begin{equation}
  \cosh(\theta) = -g(w,\tilde{w})
  = \frac{1}{\sqrt{1 - g(v_\B, v_\B)/c^2}} \; .  
\end{equation}
Thus we can express $\tilde{w}$ in terms of $w$ and the boost velocity as
\begin{align}
  \tilde{w}
  &= \cosh(\theta) \left(w + \frac{v_\B}{c}\right) \nonumber\\
  &= \frac{w + \frac{v_\B}{c}}{\sqrt{1 - g(v_\B, v_\B)/c^2}} \; .
\end{align}

Combined, we have shown that
\begin{subequations}
\begin{align}
  v_\B
  &= \frac{c \tilde{w}}{-g(w,\tilde{w})} - c w , \\
  c \tilde{w}
  &= \frac{c w + v_\B}{\sqrt{1 - g(v_\B, v_\B)/c^2}} \; .
\end{align}
\end{subequations}
With the identification $\lambda = 1/c^2$, these reproduce equations
\eqref{eq:boost_vel} for the boost velocity from the main text.

\end{document}
